\documentclass[preprint,12pt]{elsarticle}

\usepackage{amsmath}
\usepackage{colordvi}

\begin{document}
\begin{frontmatter}
  \title{Formation time dependence of femtoscopic $\pi \pi$ correlations in p+p collisions at $\sqrt{s_{NN}}$=7 TeV}
  \author[FIAS,ITP]{Gunnar Gr\"af}
  \author[FIAS,Huzhou]{Qingfeng Li}
  \author[FIAS,ITP]{Marcus Bleicher}
 
  \address[FIAS]{Frankfurt Institute for Advanced Studies, Frankfurt am Main, Germany}
  \address[ITP]{Institut f\"ur Theoretische Physik, Goethe-Universit\"at, Frankfurt am Main, Germany}
  \address[Huzhou]{School of Science, Huzhou Teachers College, Huzhou 313000, P.R. China}

  \begin{abstract}
    We investigate femtoscopic $\pi \pi$ correlations using the UrQMD approach 
    combined with a correlation afterburner. The dependence of 
    $\pi \pi$ correlations on the charged particle multiplicity and formation time in p+p collisions at
    $\sqrt{s_{NN}}$= 7 TeV is explored and compared to present ALICE data. The data allows to 
    constrain the formation time in the string fragmentation to $\tau_f \leq 0.8$ fm/c.
  \end{abstract}

\end{frontmatter}

\section{Introduction}
   With the start of the LHC physics program two years ago a tremendous amount of new data became available.
   Apart from the heavy ion data, the proton-proton (pp) program allowed to explore collective features of the strong interaction
   in high multiplicity pp events.

   It seems like in massive nucleus-nucleus collisions, a strongly interacting medium is created even in pp collisions, that 
   exhibits similar bulk properties such as space momentum correlations and collective behaviour 
   \cite{Liu:2011np,Liu:2011dk,Vogel:2010et,Werner:2010ny}. The details of these correlations 
   can be probed using Hanbury Brown-Twiss (HBT) \cite{Hanbury:1956} interferometry. While it is often argued, that the 
   particle emitting system in p+p collisions is too small to create a medium that exhibits
   bulk properties, this is different at a center of mass energy of $\sqrt{s}$= 7 TeV \cite{Werner:2011fd}.
   Here, the particle multiplicity is about the same as in nucleus-nucleus collisions, studied at
   the Relativistic Heavy Ion Collider (RHIC) in Brookhaven. For previous studies of femtoscopic correlations in
   p+p collisions at RHIC and Tevatron see \cite{:2010bw,Alexopoulos:1992iv}. This data suggests that space momentum
   correlations are developed even in pp collisions as soon,
   as high particle multiplicities are achieved. Thus, it is worthwhile studying the dependence
   of HBT observables on the event multiplicity. As the system created in p+p collisions at LHC is still small, an essential quantity
   that influences the particle freezeout radii is the formation time in flux tube fragmentation. 
   Without going into the details of the specific implementation, it is clear that the formation time sets the scale for a
   minimum value of the source lifetime - of course followed by resonance decay and rescattering. In this paper we suggest,
   that the recent LHC data on pp collisions  allows to determine the formation
time in the flux tube break-up. Results for Pb+Pb reactions and scaling studies
at the LHC within the same model can be found in \cite{Li:2011xx,Graef:2011xx}.

\section{Two-particle correlations}
   To this aim we apply two-particle correlations to extract the space-time structure of ultra-relativistic
   particle collisions. For bosons the basic equation for the correlation 
   function C(q) is
   \begin{equation}
     C({\bf q},{\bf K}) = 1 + \frac{\left | \int d^4x S(x,K) e^{iq\cdot x} \right |^2}{\left | \int d^4x S(x,K) \right |^2},
     \label{Eqn:Corr1}
   \end{equation}
   where $q = p_1 - p_2$, $K = \frac{1}{2}(p_1 + p_2)$, with $p_1$, $p_2$ being the four-momenta of the particles
   and S(x,K) is the phase-space density of the particle emitting source. Using the Pratt-Bertsch parametrization we fit the
   resulting correlation function with the functional form
   \begin{equation}
     C({\bf q},{ \bf K}) = 1+ \lambda({\bf K}) exp\left [ - \sum_{i,j=o,s,l} q_iq_jR_{ij}^2({\bf K}) \right ],
   \label{Eqn:Correlation}
   \end{equation}
   using the {\it OSL}-system. Here, {\it l} is the longitudinal direction along the 
   beam axis, {\it o} the out direction along the transverse component of {\bf K} and {\it s} the 
   side direction perpendicular to the afore mentioned directions. $R_{ij}^2$({\bf K}) are the HBT radii.
   All the radii in this paper are computed in the longitudinal comoving system (lcms).  \\

   The width of the pion pair separation in space is inversely proportional to the momentum difference {\it q} probing it.
   In the presence of flow, high p$_\perp$ particles
   that are far apart in position space are very likely to be also far apart in momentum space. This results
   in smaller HBT radii than the actual source size, because only the region of homogeneity is probed and
   not the entire source. For particles with low p$_\perp$ the effect is not as strong, resulting in a
   larger probed region, thus giving larger HBT radii.

\section{Particle formation time}
   The formation time denotes the time it takes for a hadron to be produced
from a fragmenting string. A very common model to describe such a flux tube
fragmentation is the Lund string model \cite{Andersson:1983ia}. In the Lund
model the formation time consists of the time it takes to produce an quark
antiquark pair and the time it takes for that pair to form a hadron. For the
Lund model both of these times are proportional to the transverse mass of the
created hadron and inversely proportional to the string tension. For simplicity
UrQMD uses a constant formation time of $\tau_f=0.8$ fm/c for hard collisions.
Only after the formation time particles, e.g. $\rho$-mesons, can decay or
perform subsequent scatterings with their full cross section. During their
formation time leading hadrons are allowed to interact with a reduced
cross section. Other particles are not allowed to interact at all during their
formation time. While a Heaviside function like behaviour of the cross section is
implemented in UrQMD, there are other model studies \cite{Spieles:1999pm,Gallmeister:2007an} that
investigate the influence of several scenarios of continuous changes from zero
to full cross section during the formation time. The freezeout space-time
positions of hadrons are defined as either their point of formation or the
point of last interaction, whatever occurs later in time. Since HBT probes the
freezeout distribution the extracted radii are sensitive on the value of
$\tau_f$ in small systems at high collision energies where the majority of
particles is produced from flux tube fragmentation. Although there are
many theoretical studies on the formation time
\cite{Cassing:2003sb,Arleo:2003jz,Kopeliovich:2006xy,Falter:2004uc,Bialas:1999zg}
few of them allow to put constraints on $\tau_f$ or the behaviour of the
cross section growth from experimental data.

\section{Analysis process}
   The UrQMD (v3.3p1) is used to generate the freezeout distributions of
   particles in proton-proton collisions at $\sqrt{s_{NN}}$= 7 TeV. For further information about UrQMD the reader is
   referred to \cite{Bass:1998ca,Petersen:2008dd,Li:2008qm}. After the simulation run, a correlation afterburner
   is applied to the freezeout distribution to calculate the corresponding 3D correlation function using
   Eq. \ref{Eqn:Corr1}. In this analysis, as in ALICE data, particles within a pseudorapidity interval of $|\eta|<1.2$ are taken
   into account. The analysis is done differentially for $K_\perp$ bins of 100 MeV in the region of 
   $K_\perp= 0.1 - 1.0$ GeV and also for the different event multiplicities listed in Table \ref{Tab:Multiplicities}.
   Generally the average $dN_{ch}/d\eta$ from UrQMD is 15\% smaller than the one 
   measured by the ALICE collaboration, in the same charged multiplicity classes, because we did not employ
   specific PYTHIA tunes for the present analysis.
   All the correlation functions are computed for pairs in the longitudinal comoving system (lcms).  The HBT
   radii are extracted by fitting Eq. \ref{Eqn:Correlation} to the 3-dimensional correlation functions
   over a range of $|q_i|<$ 800 MeV.

   \begin{table}
      \begin{center}
	 \begin{tabular}{c|c|c}
	   $N_{charged}^{|\eta|<1.2}$   & UrQMD $\left < dN_{ch}/d\eta \right >_{|\eta|<1.2}$ & ALICE $\left < dN_{ch}/d\eta \right >_{|\eta|<1.2}$\\
	   \hline
	   1-11   &  2.52    & 3.2  \\
	   12-16  &  5.74    & 7.4  \\
	   17-22  &  8.01    & 10.4 \\
	   23-29  &  10.72   & 13.6 \\
	   30-36  &  13.65   & 17.1 \\
	   37-44  &  16.74   & 20.2 \\
	   45-57  &  20.94   & 24.2 \\
	   58-149 &  27.57   & 31.1 \\
	 \end{tabular}
	 \caption{Table of the investigated multiplicity intervals. The first column shows the
		  interval boundaries, the second column the mean charged particle multiplicity per unit
		  of pseudorapidity ($dN_{ch}/d\eta$) from events with at least one charged particle in $|\eta| < 1.2$ from
		  UrQMD. The third column shows the same quantity from ALICE data \cite{Aamodt:2011kd}.}
	 \label{Tab:Multiplicities}
      \end{center}
   \end{table}

\section{Results}
\begin{figure}[!ht]
      \includegraphics[width=\textwidth]{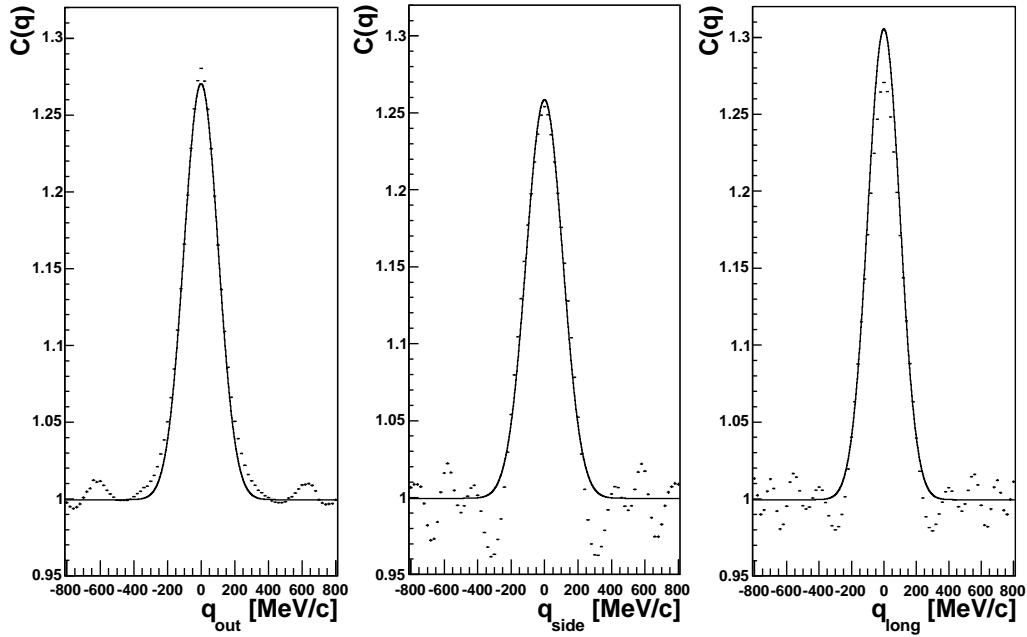}
      \caption{The dots represent projections of the 3 dimensional correlation function for 
               $K_\perp$=0.3-0.4 GeV and $N_{charged}^{|\eta|<1.2}$= 23-29. The lines represent a $\chi ^2$ fit of 
               Eq. \ref{Eqn:Correlation} to the correlation function. Both the result of the fit 
               and the correlation function are integrated over a range of $q_i=\pm 0.17$ GeV
               in the other directions for the purpose of projection.
          \label{Fig:CorrFunction}}
   \end{figure}

   In this section the results on HBT radii from the UrQMD model are compared to ALICE data 
   \cite{Aamodt:2011kd}. In Fig. \ref{Fig:CorrFunction} the projections in out, side and long 
   direction of the 3D correlation function together with a projection of the fit for 
   $K_\perp$=0.3-0.4 GeV and $N_{charged}^{|\eta|<1.2}$= 23-29 are shown as an example. It can be seen, 
   that the calculated correlation function (shown as dots) is well described by a Gaussian fit (lines).
   However, oscillations of the correlation function are present at larger {\it q}. This indicates that
   there is a non-gaussian component in the underlying separation distribution of the pion freezeout points.
   Here we will not investigate this interesting question further.  \\

\begin{figure}[!ht]
     \includegraphics[width=\textwidth]{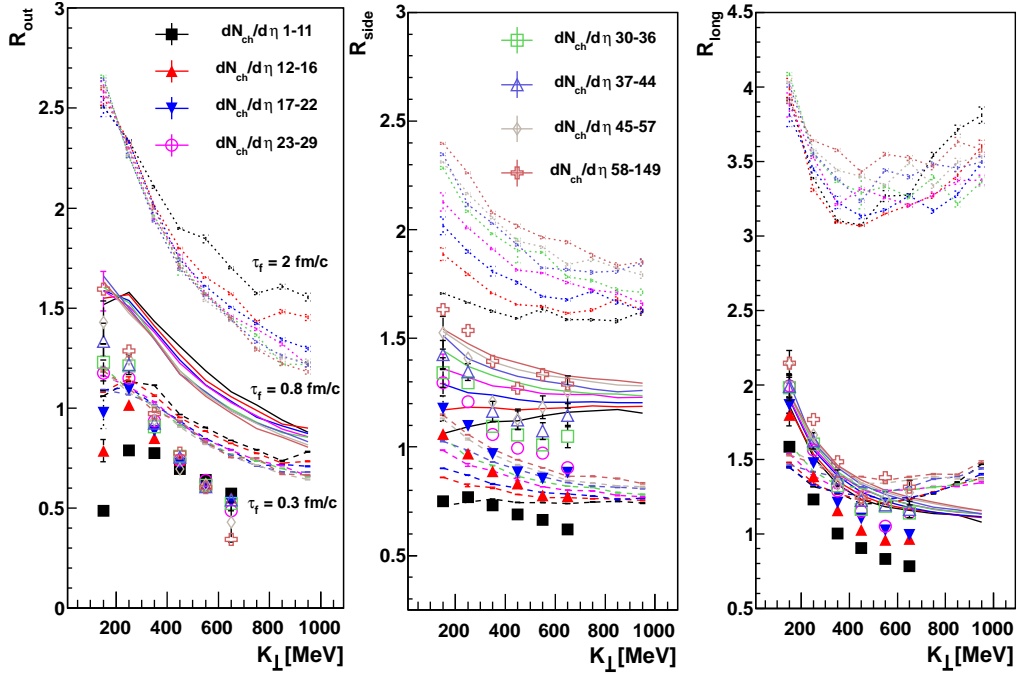}
     \caption{The lines represent HBT radii in pp collisions at $\sqrt{s}$= 7 TeV from UrQMD for different
              multiplicities and formation times. The various line styles refer to results for $\tau_f=0.3$ fm/c
              (dashed), $\tau_f=0.8$ fm/c (default - full lines) and $\tau_f=2$ fm/c (dotted). The colors 
              represent the multiplicity classes as defined in table \ref{Tab:Multiplicities}. The Points are
              data from the ALICE experiment \cite{Aamodt:2011kd}.}
     \label{Fig:RVsKt}
   \end{figure}

   The $K_\perp$ dependence of the HBT radii extracted from the UrQMD freezeout
distribution is presented in Fig. \ref{Fig:RVsKt} for the $dN_{ch}/d\eta$
classes defined in Tab. \ref{Tab:Multiplicities} in the pseudorapidity interval
$|\eta|<1.2$ in comparison to the ALICE data. The UrQMD calculations are
presented for different values of the formation time $\tau_f$ ($\tau_f$= 0.3
fm/c, dashed lines; $\tau_f$= 0.8 fm/c, full lines; $\tau_f$=2 fm/c, dotted
lines). For $\tau_f$=0.3 fm/c one obtains a good description for $R_{out}$,
while $R_{side}$ is slightly under predicted and the values for $R_{long}$ are
in line with data from ALICE. The choice $\tau_f$=0.8 fm/c leads to a slight
overestimation of $R_{out}$, however it leads to a reasonable description of
$R_{side}$ data. Also the $K_\perp$ behaviour in $R_{long}$ is much closer to
the behaviour of the data. In contrast, a formation time of $\tau_f$=2 fm/c,
leads to a drastic over estimation of the data for all radii. Although
there are discrepancies between model and data for all values of $\tau_f$, the
sensitivity on $\tau_f$ is much larger than those discrepancies. Therefore, the
present ALICE data allows to constrain the formation time to values of $\tau_f
\approx$ 0.3-0.8 fm/c.\\
   
   Let us finally discuss the overall shape of the radii as a function of
multiplicity and $K_\perp$. The $R_{side}$ radii (see Fig. \ref{Fig:RVsKt},
middle) from UrQMD and in the data are flat as a function of $K_\perp$ for low
multiplicity events. With increasing multiplicity the radii develop a decrease
towards higher $K_\perp$. This is exactly the behavior one would expect for the
development of space-momentum correlations with rising event multiplicity
\cite{Werner:2011fd}. For $R_{out}$ (Fig. \ref{Fig:RVsKt}, left) however, there
is a $K_\perp$ dependence present in all multiplicity classes. Thus, only the
development of radial flow with rising particle multiplicity seems not
sufficient to explain the $K_\perp$ dependence. Since $R_{out}$ and $R_{long}$
contain lifetime contributions of the source and $R_{side}$ does not, there
seems to be an additional non-trivial $K_\perp$ and multiplicity dependence in
the emission duration needed to explain the difference in the behaviour of
$R_{out}$, $R_{side}$ and $R_{long}$. This additional correlation might be due
to an additional momentum dependence in $\tau_f$ apart from the trivial Lorentz
boost. This would lead to a direct effect on the emission duration, because it
changes the particles production spacetime points. It would also lead to an
indirect change of the emission region, since the particle rescattering is
influenced by $\tau_f$. In this case of pp collisions the effect of the
rescattering should be negligible. Preformed hadron interactions become
important important in AA collisions \cite{Arleo:2003jz,Li:2007yd}.

\section{Summary}
   We have performed an analysis of proton-proton collisions at the top LHC
energy of $\sqrt{s} = 7 $ TeV. We have shown that the data provides rather
direct access to the particle formation times in the flux tube fragmentation.
The sensitivity to $\tau_f$ is large enough compared to the model uncertainties
to find that a value $\tau_f=0.3-0.8$ fm/c is strongly favored compared to
larger values for $\tau_f$. Values of $\tau_f \ge 2$ fm/c can be ruled out from
the present analysis.

\section{Acknowledgments}
  This work was supported by the Hessian LOEWE initiative through Helmholtz International Center for FAIR
  (HIC for FAIR). The Frankfurt Center for Scientific Computing (CSC) provided the computational resources.
  G.G. thanks the Helmholtz Research School for Quark Matter Studies (H-QM) for support. Q.L. acknowledges
  the warm hospitality of FIAS institute and thanks the financial support by the key project of the Ministry
  of Education (No. 209053), the NNSF (Nos. 10905021, 10979023), the Zhejiang Provincial NSF (No. Y6090210),
  and the Qian-Jiang Talents Project of Zhejiang Province (No. 2010R10102) of China.

\bibliographystyle{elsarticle-num}

\end{document}